\documentclass{PoS}

\title{Charged Higgs Boson Searches at the LHC via Multiple $b\bar bW^\pm$ Final States}

\ShortTitle{Charged Higgs Searches@LHC}

\author{Stefano Moretti\\
        School of Physics and Astronomy, University of Southampton,\\
	Southampton, SO17 1BJ, United Kingdom\\
        E-mail: \email{s.moretti@soton.ac.uk}}

\author{Rui Santos\\
	Centro de F\'{\i}sica Te\'{o}rica e Computacional,
	Faculdade de Ci\^{e}ncias,
	Universidade de Lisboa, \\
	Campo Grande, Edif\'{\i}cio C8 1749-016 Lisboa, Portugal\\
        E-mail: \email{rasantos@fc.ul.pt}}

\author{\speaker{Pankaj Sharma}\\
        ARC Center of Excellence for Particle Physics at the Terascale,\\ Department of Physics, University of Adelaide, 5005 Adelaide, South Australia\\
        E-mail: \email{pankaj.sharma@adelaide.edu.au}}

\abstract{We review the prospects of the
 Large Hadron Collider in accessing heavy charged Higgs boson signals in $b\bar b W^\pm$ final states, wherein the contributing channels can be $H^+\to t\bar b$, $hW^\pm$, $HW^\pm$ and $AW^\pm$. In particular, we devise a selection strategy which optimizes their global yield. We consider a 2-Higgs Doublet Model Type-II and we assume as  production mode $bg\to tH^-$ + c.c., the dominant one over the range $M_{H^\pm}\ge 480$ GeV, as dictated by $b\to s\gamma$ constraints. Possibilities  of detection are found to be significant for various Run 2 energies and luminosities. }

\FullConference{Prospects for Charged Higgs Discovery at Colliders\\
		3-6 October 2016\\
		Uppsala, Sweden}

\begin{document}

\section{Introduction}

Anticipated in the current and future runs of the Large Hadron Collider (LHC) is the discovery of a (singly-)charged Higgs boson which would be a monumental evidence of new physics  Beyond the Standard Model (BSM). Among many BSM scenarios which motivate the existence of charged scalars, 2-Higgs Doublet Models (2HDMs) are highly motivated from the perspective of Supersymmetry (SUSY) where two Higgs doublets are essential. 2HDMs provide a greater insight of the SUSY Higgs sector without including the plethora of new particles which SUSY predicts. Apart from the 2HDM Type-II (2HDM-II), which has the SUSY Yukawa structure, there can be other 2HDMS, namely, Type-I, -Y and -X depending upon how the two doublets couple to the SM fermions. In addition to a charged Higgs $H^\pm$, 2HDMs also predicts new neutral scalars,{\em viz.} the light CP-even scalar $h$, the heavy CP-even scalar $H$ and the CP-odd scalar $A$. In this work, we focus on the 2HDM-II wherein the constraints coming from $b\to s\gamma$ decays dictate the charged Higgs boson mass $M_{H^\pm}$  be larger than 480 GeV~\cite{Misiak:2015xwa}.

The LHC production of such a heavy charged Higgs  state is in association with a single top quark \cite{bg}. In this work, we study charged Higgs decaying to $W^\pm W^\mp b \bar b b$ final state originating from all bosonic decays and the top-bottom quark mode \cite{Moretti:2016jkp}. The search for charged Higgses in bosonic decays has been studied recently in \cite{Arhrib:2016wpw} and with jet substructure in \cite{Li:2016umm,Patrick:2016rtw}. Finally, charged Higgs search prospects at the LHC in a variety of channels have been detailed in \cite{Akeroyd:2016ymd}.

\section{Analysis}
\subsection{Allowed Parameter Space}

In this study, we mainly focus on the inverse alignment scenario where the heavy CP-even Higgs $H$ is the SM-like Higgs boson. The charged Higgs mass is chosen to be 500 GeV and $M_h$ to be 100 GeV. The pseudoscalar is also considered to be light with three values of its mass, namely, 100 GeV, 130 GeV and 150 GeV. This choice of masses leads to maximize the number of intermediate channels which yield a $b\bar bW^\pm$ final state via $H^\pm\to W^\pm h/H/A$ followed by the decay $h/H/A\to b\bar b$. The $H^\pm \to tb$ decay is also included in the analysis.

To obtain the allowed parameter space in the $\sin(\beta-\alpha)$ vs $\tan\beta$ plane, we fix all the masses of new scalars and make use of \textsc{ScannerS}~\cite{Coimbra:2013qq} in order to take into account  theoretical and experimental constraints. We interface \textsc{ScannerS} with 2HDMC~\cite{Eriksson:2009ws} to evaluate the decay Branching Ratios (BRs), \textsc{HiggsBounds}~\cite{Bechtle:2013wla} and \textsc{HiggsSignals}~\cite{Bechtle:2013xfa} to obtain all the constraints from collider analyses including the ones coming from LHC7 and LHC8 data.

\begin{figure}[h!]
\begin{center}
\includegraphics[scale=0.5]{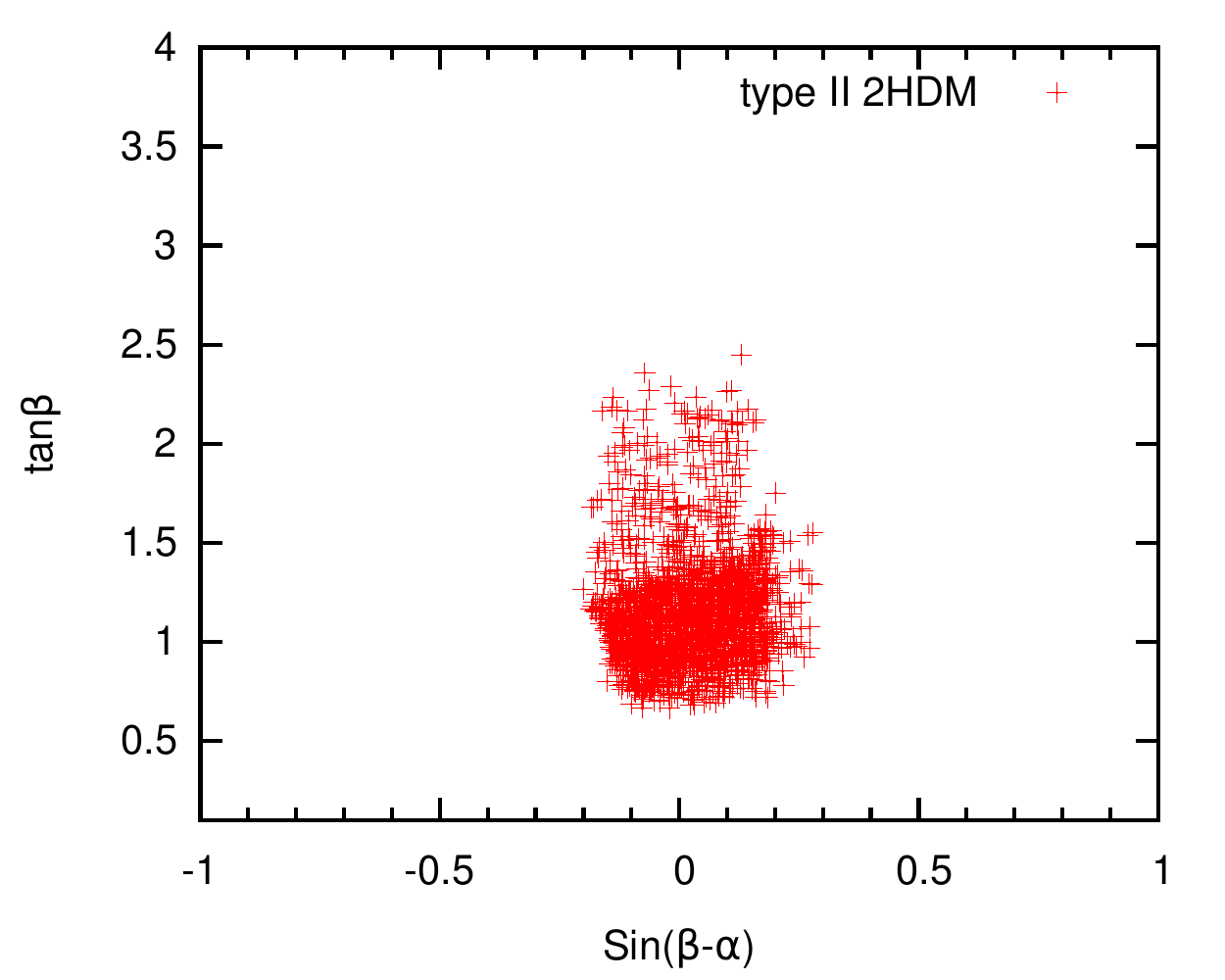}
\caption{\label{fig:points_2hdm} Allowed points on the $\sin(\beta-\alpha)$ versus $\tan\beta$ plane after the LHC Run 1.}
\end{center}
\end{figure}

In Fig.~\ref{fig:points_2hdm}, we display the points in the $\sin(\beta-\alpha)$ vs $\tan\beta$ plane allowed after  LHC Run 1 and including all theoretical constraints. From the figure, we find that  only the low $\tan\beta\lesssim 2$ and $|\sin(\beta-\alpha)|\sim 0.2$ region is favored by  current constraints. Thus, in the remainder, we fix $\tan\beta=1$ and $\sin\beta=0.1$ for our collider simulation. 

\subsection{Decays of Charged Higgs Bosons}

\begin{figure}[h!]
\begin{center}
\includegraphics[scale=0.5]{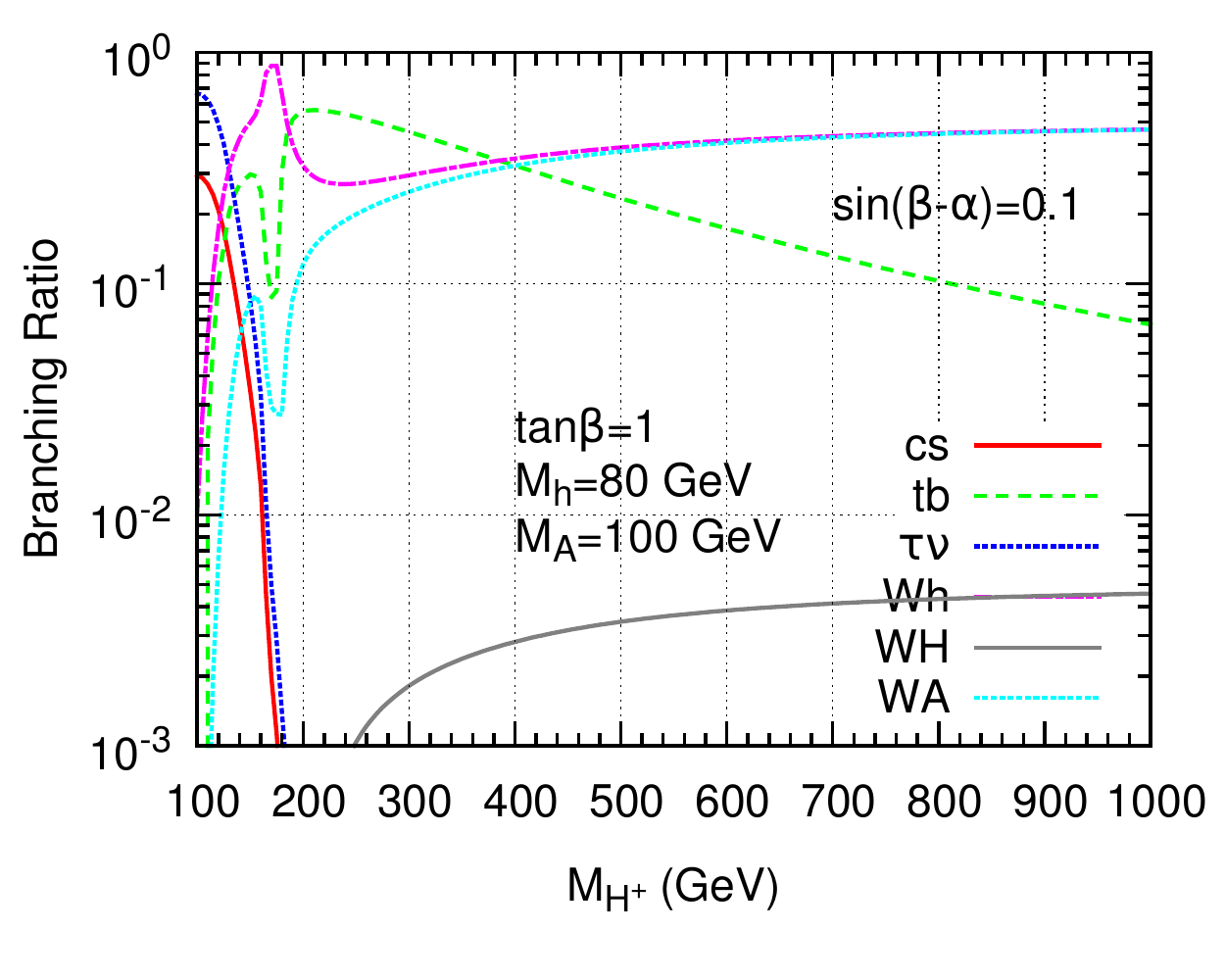}
\includegraphics[scale=0.5]{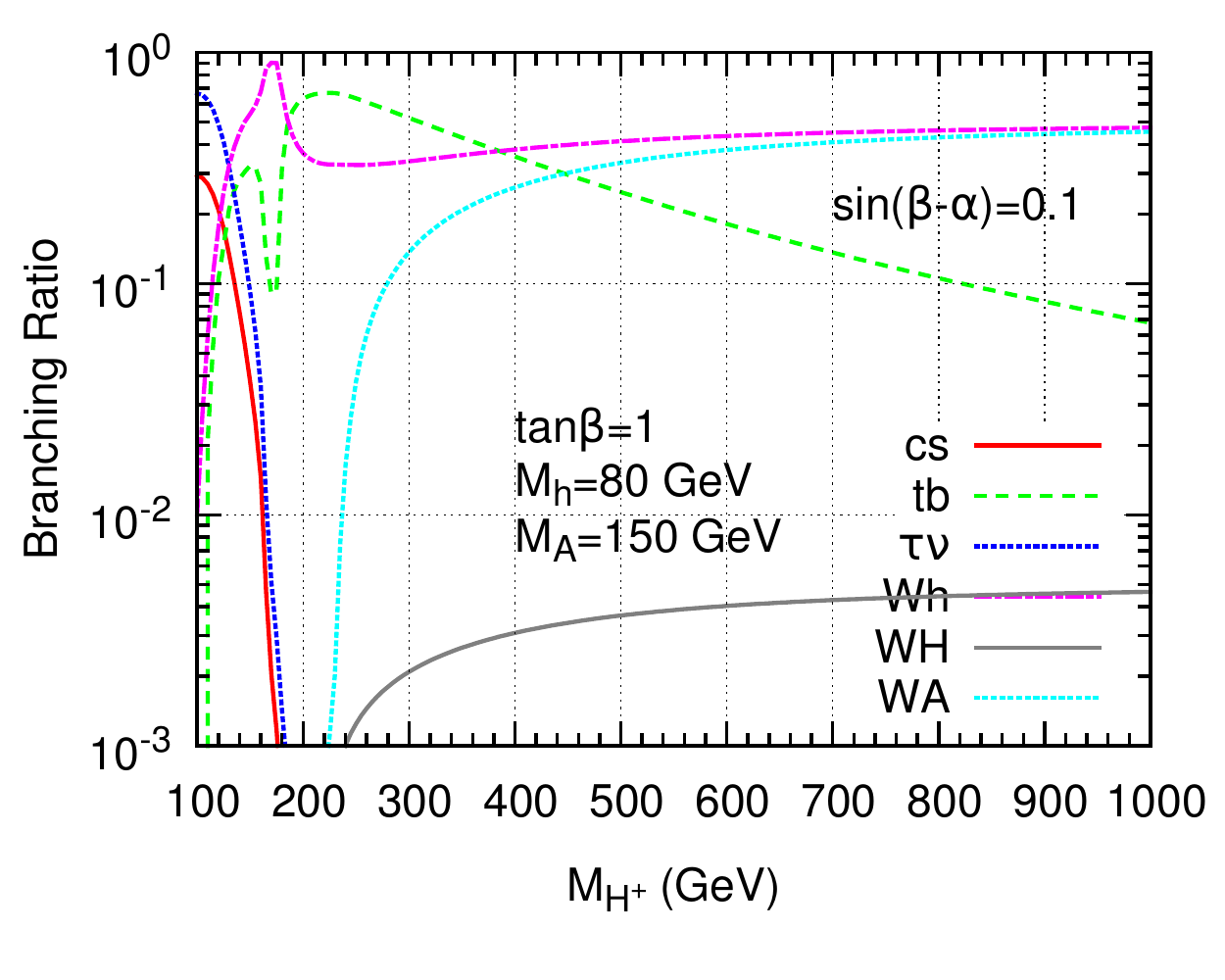}
\caption{\label{fig:br_2hdmII} BRs for a 2HDM-II charged Higgs boson with and $M_A = 100$ GeV (left),  and $M_A = 150$ GeV (right). The remaining parameters are fixed to $M_H = 125$ GeV, $M_h = 80$ GeV, $\tan\beta=1$ and $\sin (\beta-\alpha) = 0.1$.}
\end{center}
\end{figure}

In the previous section, we discussed the constraints on the parameter space from current LHC data and fixed  parameters such as $\tan\beta$, $\sin(\beta-\alpha)$ and other masses of scalars. For such a parameter space and chosen masses, several decay channels open up. In Fig.~\ref{fig:br_2hdmII}, we present the BRs of various decay channels of the charged Higgs state for two values of pseudoscalar mass, $M_A=100$ GeV (left) and 150 GeV (right),  respectively. We find that, for a chosen benchmark point with $M_{H^\pm}=500$ GeV, the BRs of the decay modes $W^\pm h$ and $W^\pm A$ are about 38\%-40\%, respectively, while for the $tb$ mode  is around 8\%-10\%. We find, in general, that, as soon as the $H^\pm$ bosonic decays are allowed, they  become dominant over the entire parameter space. In the following, we  consider all decay channels and their possible interference effects in numerical simulation.

\subsection{Signal and Backgrounds}
The dominant production of a heavy charged Higgs boson is in associated production with a single top quark, i.e., $pp\to t H^- + c.c.$, at the LHC with total cross section at around 900 fb for $M_{H^\pm}=500$ GeV in the 2HDM-II at leading order. The charged Higgs decays to the $W^\pm b\bar b$ final state originate from four decay channels, the $W^\pm h/H/A$ and the $tb$ modes. Thus the final signal includes two $W^\pm$ bosons and 3 $b$-jets. We consider one of the $W^\pm$ bosons to decay leptonically and other hadronically. 

The irreducible background to our signal process comes from $W^+W^- b\bar b b$ processes which include SM single top production as well as $t\bar t b$ processes with a total cross section of about 9 pb. The dominant contribution to background originates from $W^+W^- b\bar b j$ processes that include  top pair production and decay. Another background considered is the $W^+W^- b j j$ process, which can be suppressed efficiently with the requirement of 3 $b$-jets in the  event sample.

\subsection{Simulation Setup}
We generate the parton level signal and background events at leading order and then pass these events through parton showering and hadronization, then, they are passed through  a fast detector simulation (see Ref.~\cite{Moretti:2016jkp}). 
Below we briefly discuss our identification and selection criteria.
\begin{itemize}
\item {\bf Identification cuts}
 \begin{enumerate}
  \item Events must have at least 1 lepton ($e$ or $\mu$), 3 $b$-jets and at least 2 light jets,
  \item All leptons and jets must satisfy: $p_{Tj,\ell}>20~ \mbox{GeV},~ |\eta_{j,\ell}|<2.5,$
  \item All pairs of objects must be well separated from each other,
  $$\Delta R_{jj,jb,bb,\ell j,\ell b}\geq 0.4~~ \mbox{where}~~\Delta R=\sqrt{(\Delta \phi)^2+(\Delta \eta)^2}.$$
 \end{enumerate}
 
\item {\bf Selection requirements} 

When an event satisfies all above requirements, it is further processed for signal reconstruction and background reduction as follows.
\begin{figure}[h!]
\begin{center}
\includegraphics[scale=0.225]{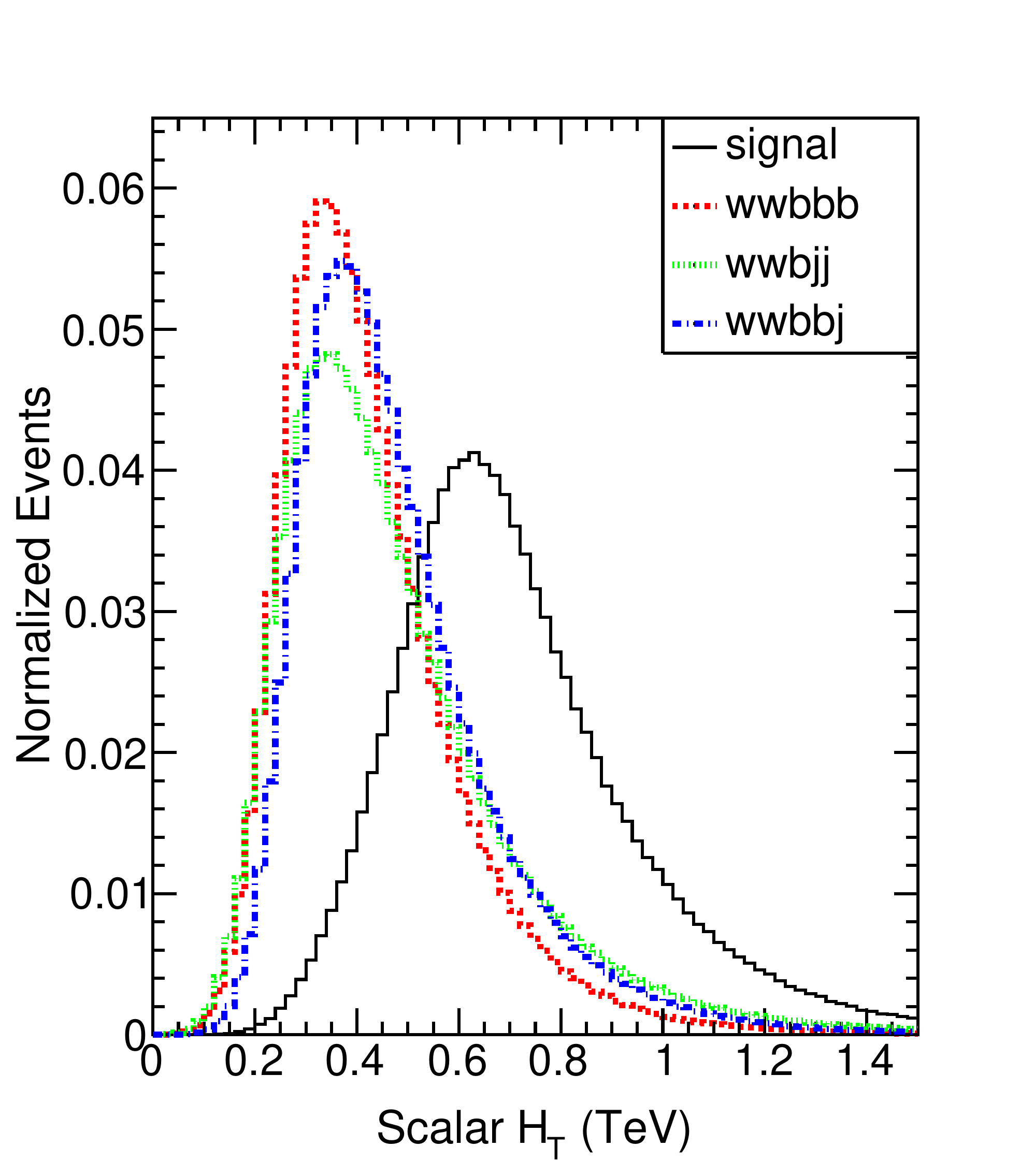}
\includegraphics[scale=0.225]{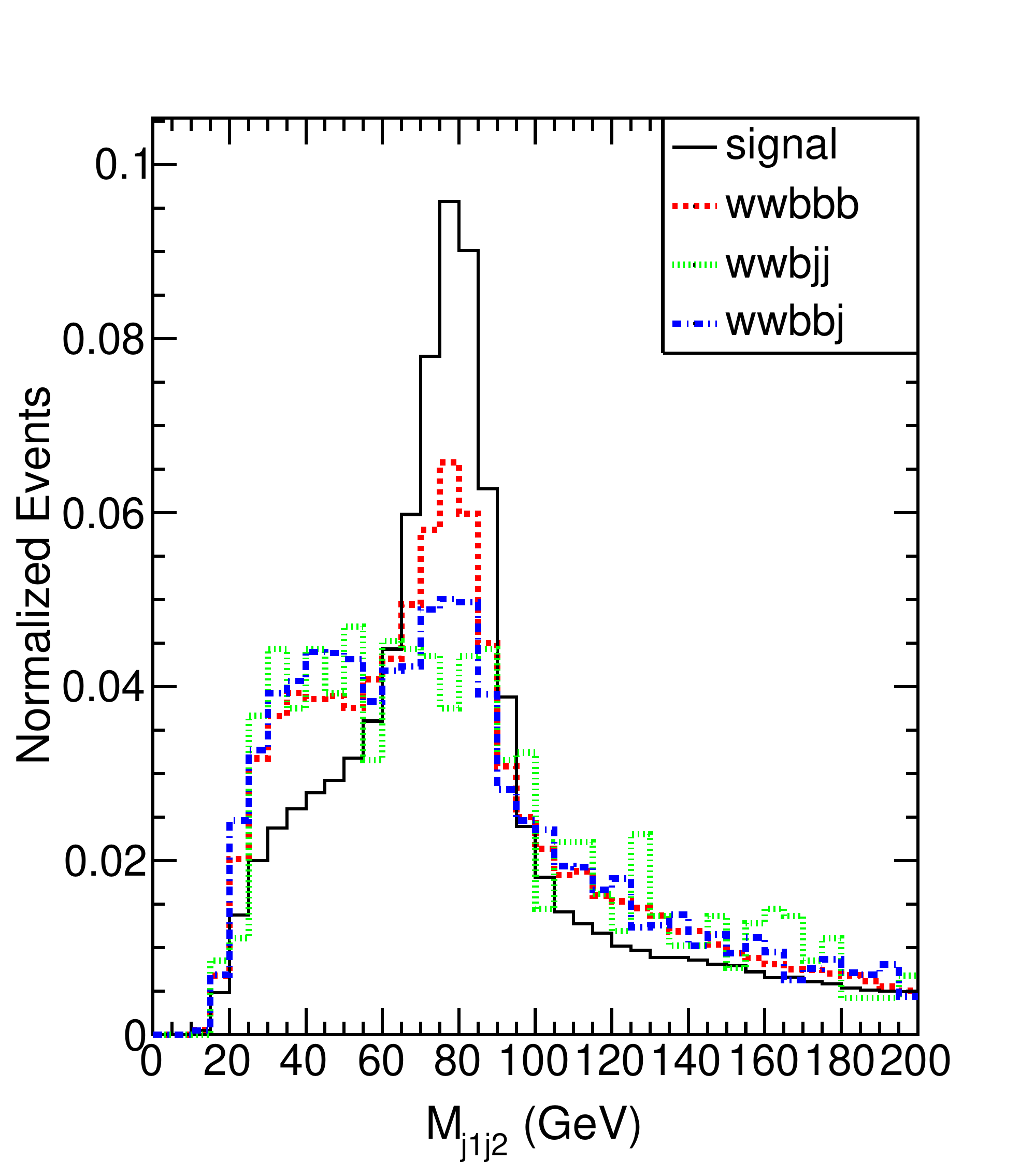}
\caption{\label{fig:HT} Scalar sum of $p_T$'s ($H_T$) distribution (left) and invariant mass $M_{j_1 j_2}$ of two light jets with minimum $\Delta_R$ (right) for signal and backgrounds.}
\end{center}
\end{figure}

\begin{enumerate}
\item {\bf $b$ tagging efficiency}: the $b$ tagging efficiency is chosen according to following rule: $\epsilon_{\eta}\tanh(0.03\;p_T-0.4),$ where $\epsilon_\eta=0.7$ for $|\eta|\leq 1.2$ and 0.6 for $1.2\leq|\eta|\leq 2.5$. The $c$-jet faking probability as a $b$-jet is obtained from the same expression but now with $\epsilon_\eta=0.2$ for $|\eta|\leq 1.2$ and $\epsilon_\eta=0.1$ for $1.2\leq|\eta|\leq 2.5$.

\item {\bf Cut on $H_T$}:  a useful variable is the scalar sum of the $p_T$'s of all the visible particles in the final state, $H_T=p_T^{\ell^\pm}+\sum_j p_T^j$.
Fig.~\ref{fig:HT} (left panel) shows the $H_T$ distributions for the signal and backgrounds. The signal events include a heavy particle which produces high-$p_T$ decay products and thus has a peak at large $H_T$. A cut on $H_T > 500$ GeV reduces the $WWbbj$ and $WWbbb$ backgrounds to 36\% and 27\% of their initial values, respectively, while the signal events are only decreased to 87\% of their initial values.

\item {\bf Hadronic $W^\pm$ candidate}: a heavy charged Higgs state leads to a highly boosted $W^\pm$ and $b\bar b$ pairs leading to their decay products to be closely spaced. We reconstruct the hadronic $W^{\pm}$ from the jets with $\Delta R_{\mathrm{min}}$ as shown in the right panel of Fig.~\ref{fig:HT}.

\item {\bf Leptonic $W^\pm$:}  a leptonically decaying $W^\pm$ is reconstructed using the information about the missing transverse momentum and imposing the invariant mass constraint $M_{l\nu}^2 = M_{W^\pm}^2$. Using the momenta of the reconstructed neutrino and  lepton, the momentum of the leptonic $W^\pm$ can be obtained.

\begin{figure}[h!]
\begin{center}
\includegraphics[scale=0.225]{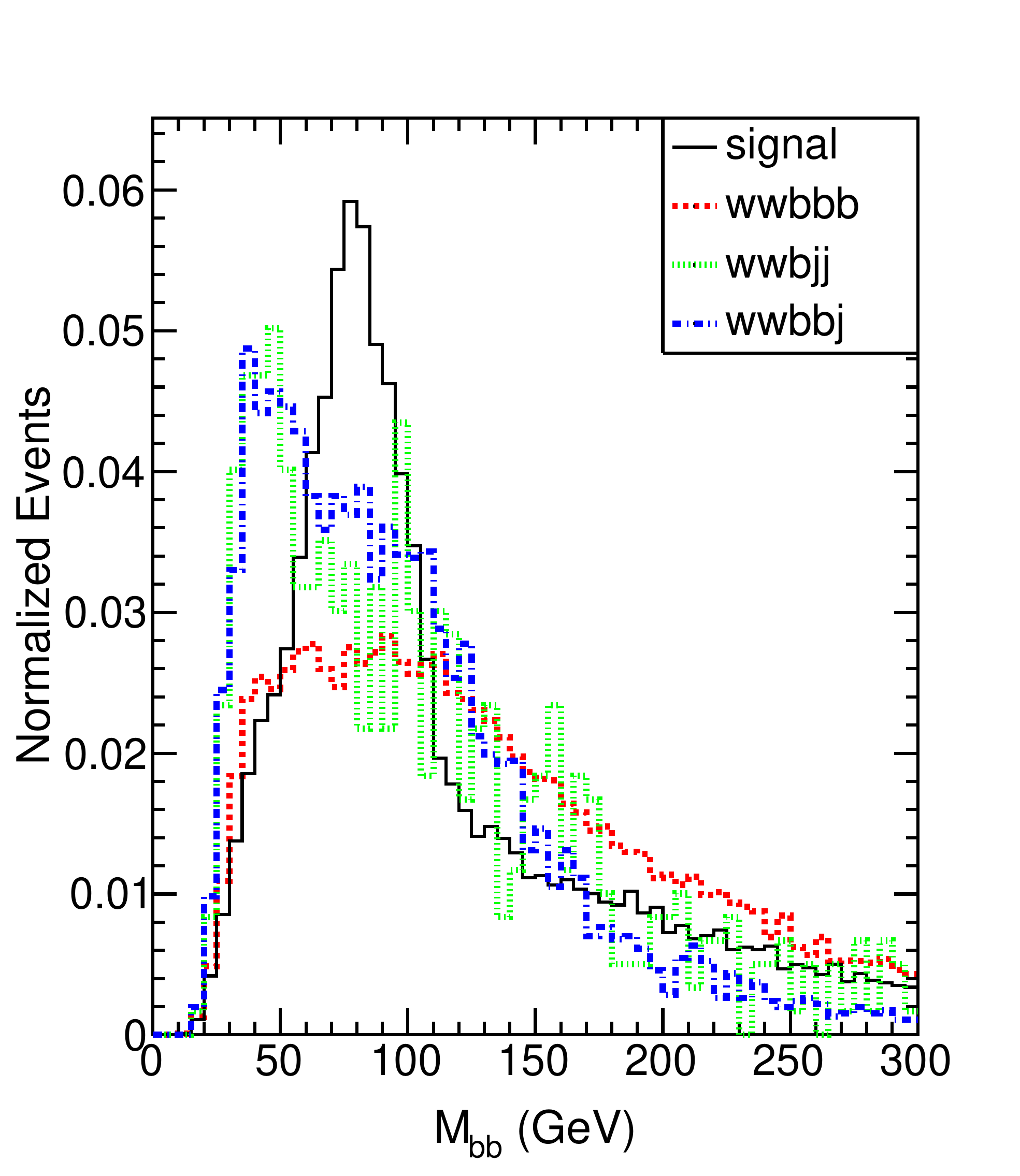}\hspace{-0.25cm}
\includegraphics[scale=0.225]{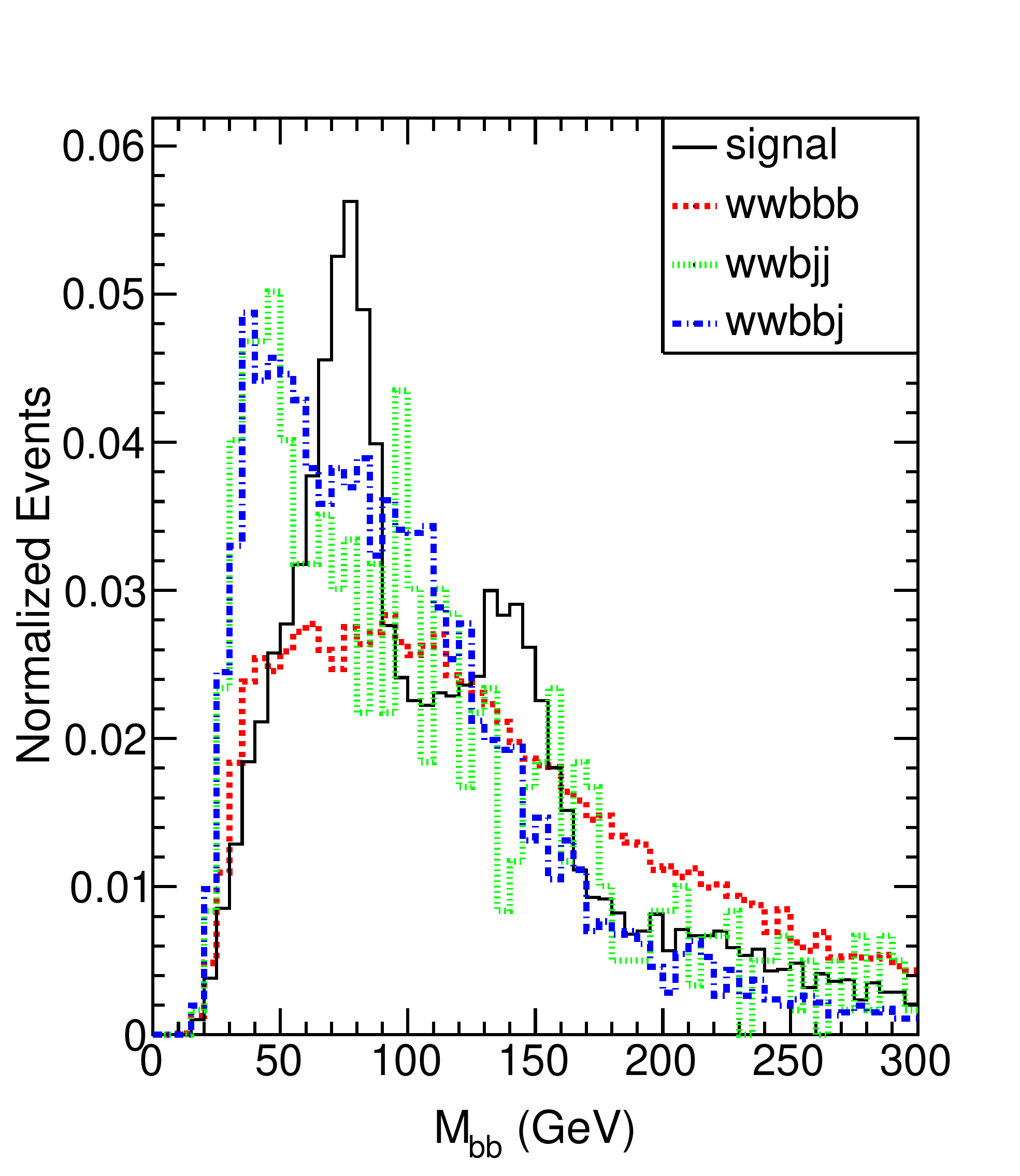}
\caption{\label{fig:INV_Mbb} Invariant mass ($M_{bb}$) of 2 $b$-jets with $\Delta R_{\rm min}$ for $M_A=$ 100 GeV (left), 130 GeV (middle) and 150 GeV (right)  for the signal and backgrounds. As discussed in the text, the 2 $b$-jets with  minimum $\Delta R$ are chosen to reconstruct $h$ and $A$.}
\end{center}
\end{figure}
 
\end{enumerate}
\end{itemize}

\begin{figure}[h!]
\begin{center}
\includegraphics[scale=0.225]{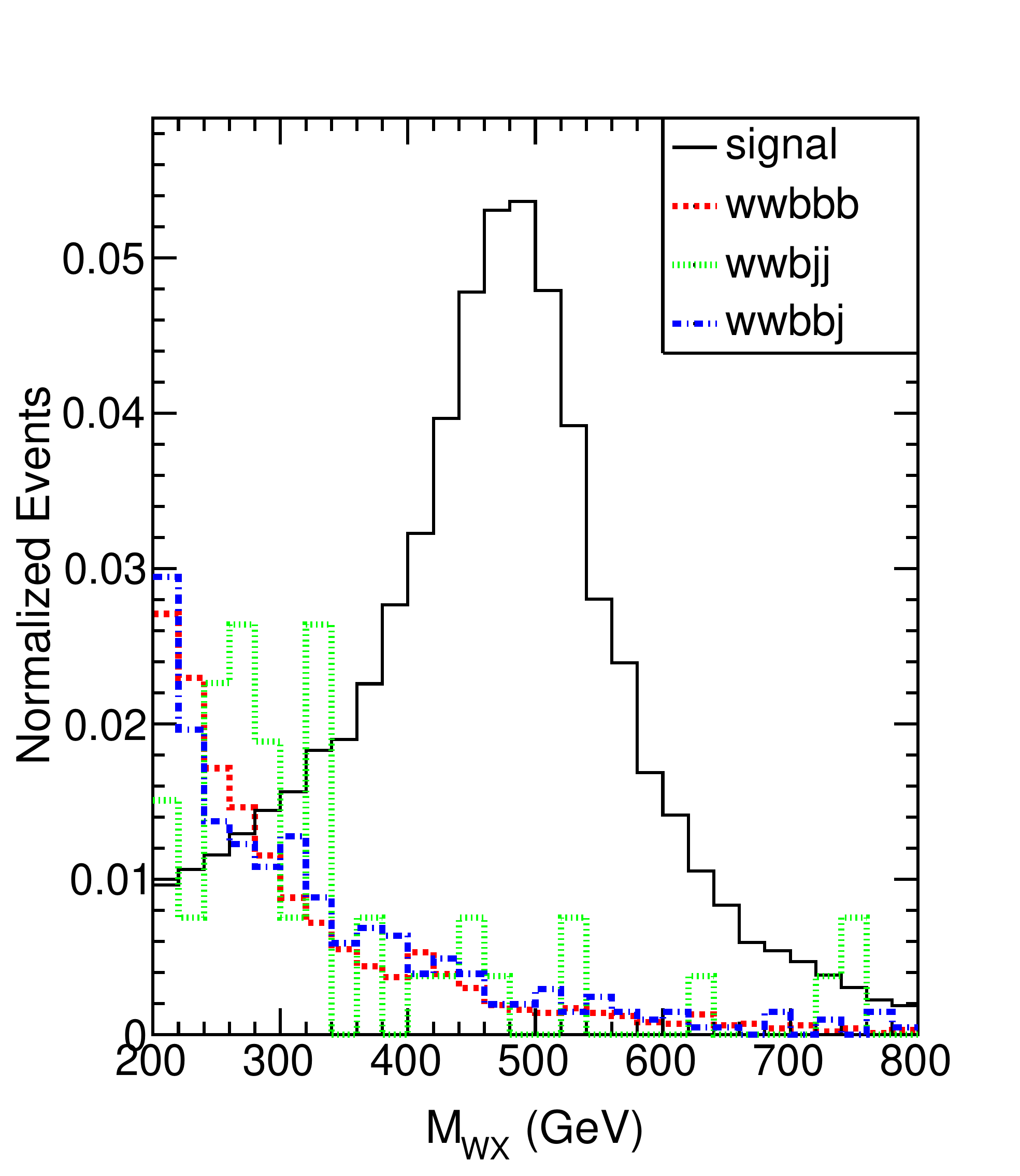}
\includegraphics[scale=0.225]{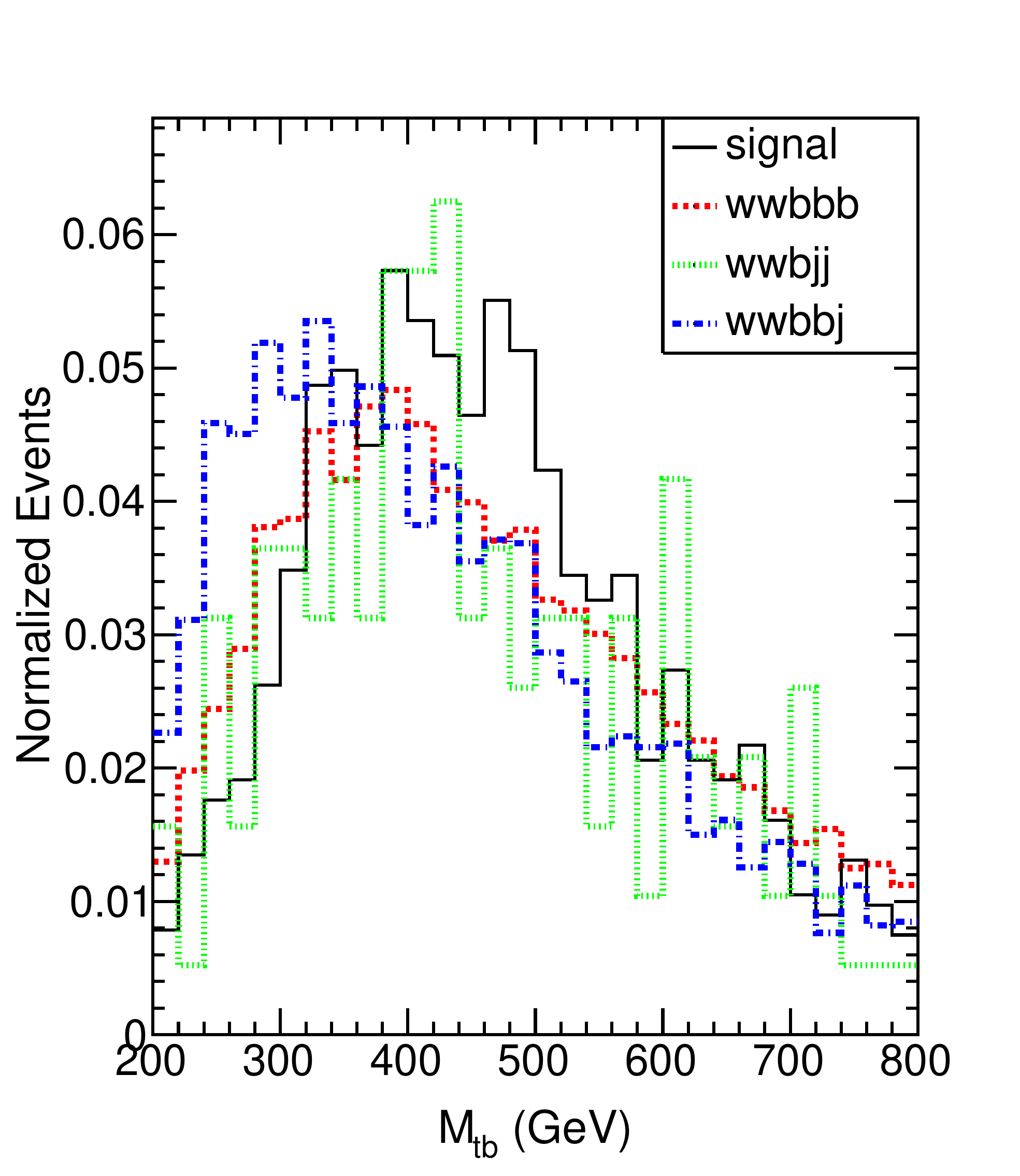}
\caption{\label{fig:INV_MWh} Invariant mass ($M_{WX}$) of the other $W^\pm$ and of the reconstructed $h,~A$ state (left) and of the reconstructed top and the remaining $b$-jet $M_{tb}$ for the signal and backgrounds. }
\end{center}
\end{figure}

After reconstructing the $W^\pm W^\mp b\bar b b$ final state, we now proceed to extract individual signal channels by applying additional sets of cuts. For the case of the $W^\pm h$ and $W^\pm A$ models, we first reconstruct the neutral Higgs mass by finding the pairs of $b$-jets with minimum $\Delta_R$ which have been shown in Fig.~\ref{fig:INV_Mbb} for $M_A=100$ GeV and $M_A=150$ GeV. One can see two peak in right panel corresponding to $h$ and $A$ while in the left panel the two peaks are joined together displaying a large width in $M_{b\bar b}$ distribution. Finally, we combine the $h$ or $A$ momentum with the $W^\pm$ boson one to reconstruct the charged Higgs mass as shown Fig.~\ref{fig:INV_MWh} (left panel).

From the remaining events, we first reconstruct the other top from thethe  reconstructed $W^\pm$ and remaining $b$-jets which give the best fit to the  top quark mass. Then we combine one of the reconstructed top's with the remaining $b$-jets to reconstruct the charged Higgs mass. The one which gives a better reconstruction is kept. We display the invariant mass $M_{tb}$ in the right panel of Fig.~\ref{fig:INV_MWh} for signal and backgrounds.

\section{Conclusions}
We have chosen a heavy charged Higgs scenario where all possible decay channels are kept open. Since they all contribute to the most relevant signature, which is $W^\pm W^\mp b\bar b$, we have considered the simultaneous contribution of the different intermediate states $W^\pm h$, $W^\pm A$, $W^\pm H$ (which is however subleading as we have taken $H$ to be SM-like) and $tb$. The signal mode is associated production of a top-quark and a charged Higgs boson. We also evaluate the efficiency through which each signal can be extracted from the LHC data. From the final number of signal and background events after applying all set of cuts, we evaluate the signal-to-background significance for the global $W^\pm X$ signal to be over 5$\sigma$ with just 100 fb $^{-1}$ of integrated luminosity. In constrast, to attain a similar significance in the $tb$ mode, we find that we will need a full 3000 fb$^{-1}$ of data set which is projected to be accumulated after the LHC14 run is completed \cite{Gianotti:2002xx}.\vskip0.35cm

\noindent{\bf Acknowledgements:}~SM is supported in part through the NExT Institute. SM and RS are supported by the grant H2020-MSCA-RISE-2014 no. 645722 (NonMinimalHiggs). PS is supported by the University of Adelaide and the Australian Research Council through the ARC Center of Excellence for Particle Physics (CoEPP) (grant no. \ CE110001004).

\end{document}